\documentclass[pra,twocolumn,amsmath,amssymb,superscriptaddress]{revtex4}

\usepackage{graphicx}
\usepackage{bm}
\usepackage{hyperref}

\newcommand{\fig}[4][t]{\begin{figure}[#1]\begin{center}\includegraphics[scale=#2]{#3}\vspace{-0.25 cm}\caption{#4}\label{fig:#3}\end{center}\end{figure}}
\newcommand{\ket}[1]{\ensuremath{\left|{#1}\right\rangle}}
\newcommand{\Rb}[1]{\ensuremath{^{#1}}\textrm{Rb}}
\newcommand{\hf}[2]{\ket{#1, \; #2}}
\newcommand{\units}[1]{\textrm{\thinspace #1}}

\begin{document}

\title{Cold atom gravimetry with a Bose-Einstein condensate}

\author{J. E. Debs}
\email{john.debs@anu.edu.au}
\homepage{http://atomlaser.anu.edu.au}
\homepage{http://www.acqao.org}
\author{P. A. Altin}
\author{T. H. Barter}
\author{D. D\"oring}
\author{G. R. Dennis}
\author{G. McDonald}
\affiliation{Australian Centre for Quantum Atom Optics and Department of Quantum Science, \\ 
the Australian National University, Canberra 0200, Australia}
\author{R. P. Anderson}
\affiliation{School of Physics, Monash University, Melbourne 3800, Australia}
\author{J. D. Close}
\author{N. P. Robins}
\affiliation{Australian Centre for Quantum Atom Optics and Department of Quantum Science, \\ 
the Australian National University, Canberra 0200, Australia}

\date{\today}

\begin{abstract}

We present a cold atom gravimeter operating with a sample of Bose-condensed $^{87}$Rb atoms. Using a Mach-Zehnder configuration with the two arms separated by a two-photon Bragg transition, we observe interference fringes with a visibility of ($83\pm6$)\% at $T=3$\,ms. We exploit large momentum transfer (LMT) beam splitting to increase the enclosed space-time area of the interferometer using higher-order Bragg transitions and Bloch oscillations. We also compare fringes from condensed and thermal sources, and observe a reduced visibility of ($58\pm4$)\% for the thermal source. We suspect the loss in visibility is caused partly by wavefront aberrations, to which the thermal source is more susceptible due to its larger transverse momentum spread. Finally, we discuss briefly the potential advantages of using a coherent atomic source for LMT, and present a simple mean-field model to demonstrate that with currently available experimental parameters, interaction-induced dephasing will not limit the sensitivity of inertial measurements using freely-falling, coherent atomic sources.

\end{abstract}

\maketitle

\section{INTRODUCTION}
In principle, a light bulb and laser that have the same photon flux will yield the same precision in many shot noise limited optical measurements. In practice, it is often the classical properties of optical lasers \textendash\ the brightness, coherence, and low phase and amplitude noise \textendash\ that enable a shot noise limited optical measurement at high flux. To date, inertial measurements using atom interferometers have primarily utilised cold thermal sources \cite{Kasevich:1992aa,Canuel:2006ab,Muller:2009aa}. It is therefore of interest to investigate whether coherent, high brightness atomic sources such as Bose-Einstein condensates (BECs) will have similar advantages in such measurements. To date, the use of BEC has been largely neglected by the precision measurement community. One concern is that the comparatively high atom density will lead to interaction-induced dephasing, thereby limiting precision \cite{Castin:1997aa,Fattori:2008aa}.

Here we present results on a gravimeter based on atomic interference \cite{Peters:2001aa,Muller:2008ab} of Bose-condensed $^{87}$Rb, and compare its performance to that achieved with a cold thermal sample in the same system. We observe an increase in fringe visibility when using a condensed source instead of a thermal one in an equivalent setup. With one of the most promising avenues for increasing sensitivity being large momentum transfer (LMT) beamsplitting \cite{Denschlag:2002aa,Muller:2008aa,Muller:2009ab,Clade:2009aa}, we use LMT to increase our sensitivity to gravity, and maintain good fringe visibility for both Bragg- and Bloch-based LMT. Furthermore, using a simple model we demonstrate that dephasing in an expanded BEC will not limit the precision of inertial measurements.

%%%%%%

\section{BACKGROUND AND METHODS}

The operating principle of a gravimeter based on atomic interference has been described elsewhere \cite{Kasevich:1992aa,Peters:2001aa}. Our gravimeter uses $n$th-order Bragg transitions \cite{Giltner:1995aa,Kozuma:1999ab,Muller:2008aa} as our atom-optic beamsplitters ($\pi/2$ pulses) and mirrors ($\pi$ pulses) in a Mach-Zehnder ($\pi/2 - \pi - \pi/2$)  configuration. These couple vertical momentum states separated by $2n\hbar k$, where $k=|\vec{k}|$ is the wavenumber of the light and $n$ an integer. For uniform acceleration the atomic phase evolution of each arm is identical, and the only interferometric phase contribution is from the atom-light interaction \cite{Kasevich:1992aa}:
\begin{eqnarray}
\Phi = n(\phi_1-2\phi_2+\phi_3) = 2n \, \vec{k}\cdot\vec{g}T^2 
\label{int_phase}
\end{eqnarray}
where $\phi_i$ is the optical phase of the $i$th Bragg pulse, and $T$ is the time between pulses. Scanning $\Phi$ results in fringes $P=\frac{1}{2}(A+V\cos{\Phi})$ in the relative population in state $|p_0+2n\hbar k\rangle$, where we define $V$ as the visibility \cite{note2} and $A$ the fringe offset. One infers $\Phi$ by operating at mid-fringe, where the change in $P$ for a given phase shift is maximal. For small shifts $\Delta\Phi$ one obtains a signal of $\Delta P = \frac{V}{2}\Delta\Phi = Vn\vec{k}\cdot \Delta \vec{g}T^2$. High signal gain thus requires having a high visibility and a large space-time enclosed area ($\propto 2nkT^2$). BEC interferometers have already been shown to exhibit fringe visibility close to 100\% \cite{Torii:2000aa}. The evolution time $T$ is typically limited due to practical considerations such as the fall distance of the atoms \cite{Peters:2001aa}, and low-frequency phase-noise sensitivity \cite{Le-Gouet:2008aa}. Increasing the space-time area using large-momentum-transfer (LMT) beamsplitting can be achieved using higher-order ($n\geq2$) Bragg transitions \cite{Muller:2008aa} or Bloch oscillations \cite{Denschlag:2002aa,Muller:2009ab,Clade:2009aa}.

LMT techniques require the momentum width of the source along $\vec{k}$ to satisfy $\Delta p \ll \hbar k$, so that the entire cloud may be coupled to a single momentum state. As suggested in \cite{Hughes:2009aa}, this requirement on $\Delta p$ becomes more stringent for increasing $n$ in order to maintain a high LMT efficiency. This can be understood by considering the dispersion relation, $\Delta E_n / \Delta p = p_n/m = 2n\hbar k/m$, where $E_n$ is the energy of the $n$th Bragg resonance. This shows that a cloud of given momentum width $\Delta p$ has a greater spread in energy at higher momentum. Thus, for a Bragg transition resonant with the centre of the cloud, there will be greater spread in detuning from resonance across the cloud for higher order transitions.

Using Bragg spectroscopy \cite{Stenger:1999aa}, we measure the 1-$\sigma$ momentum width of our cloud to be $\Delta p = 0.14\hbar k$ after 12\,ms of ballistic expansion (see Fig. \ref{fig:meanfield}c). In contrast, even the coldest reported thermal source at a temperature of $150\,$nK has $\Delta p = 0.87\hbar k$ \cite{Muller:2008ab}. More typical thermal sources at $\sim 2\,\mu$K require velocity selection at the cost of atom number to achieve the required width. One can further reduce the momentum width of a condensate by manipulating trap parameters \cite{Kozuma:1999ab} or tuning the atomic interactions using a Feshbach resonance. BECs also have a transverse spatial width over an order of magnitude lower than a typical thermal source. Coupled with a low transverse momentum width, this leads to relaxed constraints on spatial wavefront aberrations and intensity gradients in the LMT lasers.

Our production of BEC is described in \cite{Altin:2010ab}. Briefly, we produce pure \Rb{87} \hf{F=1}{m_F=-1} condensates with up to $2\times 10^6$ atoms in a crossed optical-dipole trap with $\omega_{x,y,z} = 2\pi\times(50,57,28)\units{Hz}$ by evaporation in a magnetic and an optical trap. We then switch off the trap suddenly, allowing the cloud to expand and fall for up to 35\,ms before probing the atoms using standard absorption imaging. We can transfer the atoms to the magnetically-insensitive \ket{m_F=0} state using a Landau-Zener rf sweep after the BEC is formed; however, we presently find this step unnecessary as we observe no effect on fringe visibility, signal-to-noise ratio or our measurement of $g$. To produce a thermal cloud for comparison, we terminate the magnetic trap evaporation earlier, loading fewer atoms into the optical trap. The final evaporation then results in a phase space density just below that required for condensation at 100\,nK. After releasing the cloud we wait 12\,ms before initiating our interferometer cycle. 

Building a gravimeter using a BEC source allows spatially resolved imaging of the different momentum states, eliminating the need for state labelling via Raman transitions. This cancels systematic shifts \cite{Le-Gouet:2008aa} and greatly simplifies our laser system. We derive the two phase-locked optical frequencies required to drive Bragg transitions from an amplified external-cavity diode laser giving 1.3\,W of light red detuned 90\,GHz from the $\ket{5S_{1/2}, F=1} \rightarrow \ket{5P_{3/2}, F^\prime=2}$ transition. The light is split through two 80\,MHz acousto-optic modulators (AOMs) driven by a direct digital synthesizer (DDS) referenced to a rubidium frequency standard. This allows us to produce the arbitrary pulse shapes and coherent frequency sweeps required for the experiment. The first order of each AOM is combined on a polarizing beamsplitter and coupled into a polarization-maintaining fiber, resulting in 150\,mW in a collimated 3\,mm beam directed vertically through our science cell. The beam passes through a $\lambda/4$ waveplate before being retro-reflected by a mirror mounted on a multi-layer passive vibration isolation system. This scheme also generates a second pair of Bragg frequencies, but as we allow 12\,ms of free fall to reach ballistic expansion before initiating the interferometer sequence, this pair is Doppler shifted well off resonance. We measure the relative phase noise of the Bragg laser system to be negligible compared to that introduced by our retro-mirror.

The Bragg resonance condition is given by $\delta_n = 4n\omega_r$, where $\delta_n$ is the frequency difference of the Bragg beams for an $n$th order transition, and $\omega_r = \hbar k^2/2m$ is the single-photon recoil frequency. To operate in the Bragg regime and address the entire cloud, we use Gaussian-shaped pulse envelopes \cite{Muller:2008ac} and choose our pulse length $\tau$ to satisfy $\frac{\Delta\vec{p}}{m}\cdot\vec{k} < \tau^{-1} < \omega_r$, ensuring minimal loss to adjacent momentum states for a given $n$. For $n=1$, we are able to maximise our $\pi$-pulse efficiency to 95\% in this way. However, we find that a $300\,\mu$s velocity selection pulse is required in order to achieve 93\% efficiency for $n=3$, an observation that supports the argument given earlier that the requirement on $\Delta p$ becomes more stringent with $n$. Due to the size of our science cell, we are limited to an interrogation time of $T = 5\,$ms. After several milliseconds of further separation following the final $\pi/2$ beamsplitter, an absorption image is taken to determine the number of atoms in each momentum state.

The freely-falling atoms experience a time dependent Doppler shift $\delta_d(t) = 2\pi \alpha_0 t$ where $\alpha_0 = \frac{1}{\pi} \vec{k}\cdot\vec{g}$ is a frequency chirp. This modifies the Bragg resonance condition in the laboratory frame to $\delta_n(t) = 4n\omega_r+2\vec{k}\cdot\vec{g}t$. We compensate this by sweeping $\delta$ at a rate $\alpha \simeq 25.1\,$MHz/s, determined by local $\vec{g}$ near our lab in Canberra, Australia \cite{Amalvict:2001aa}. The interferometer phase then becomes $\Phi = n(2\vec{k}\cdot\vec{g}T^2 - 2\pi\alpha T^2)$. By scanning the sweep rate $\alpha$, we record interference fringes with a period of $1/nT^2$. 

\section{GRAVIMETRY WITH BEC}

%%%%
\fig{0.6}{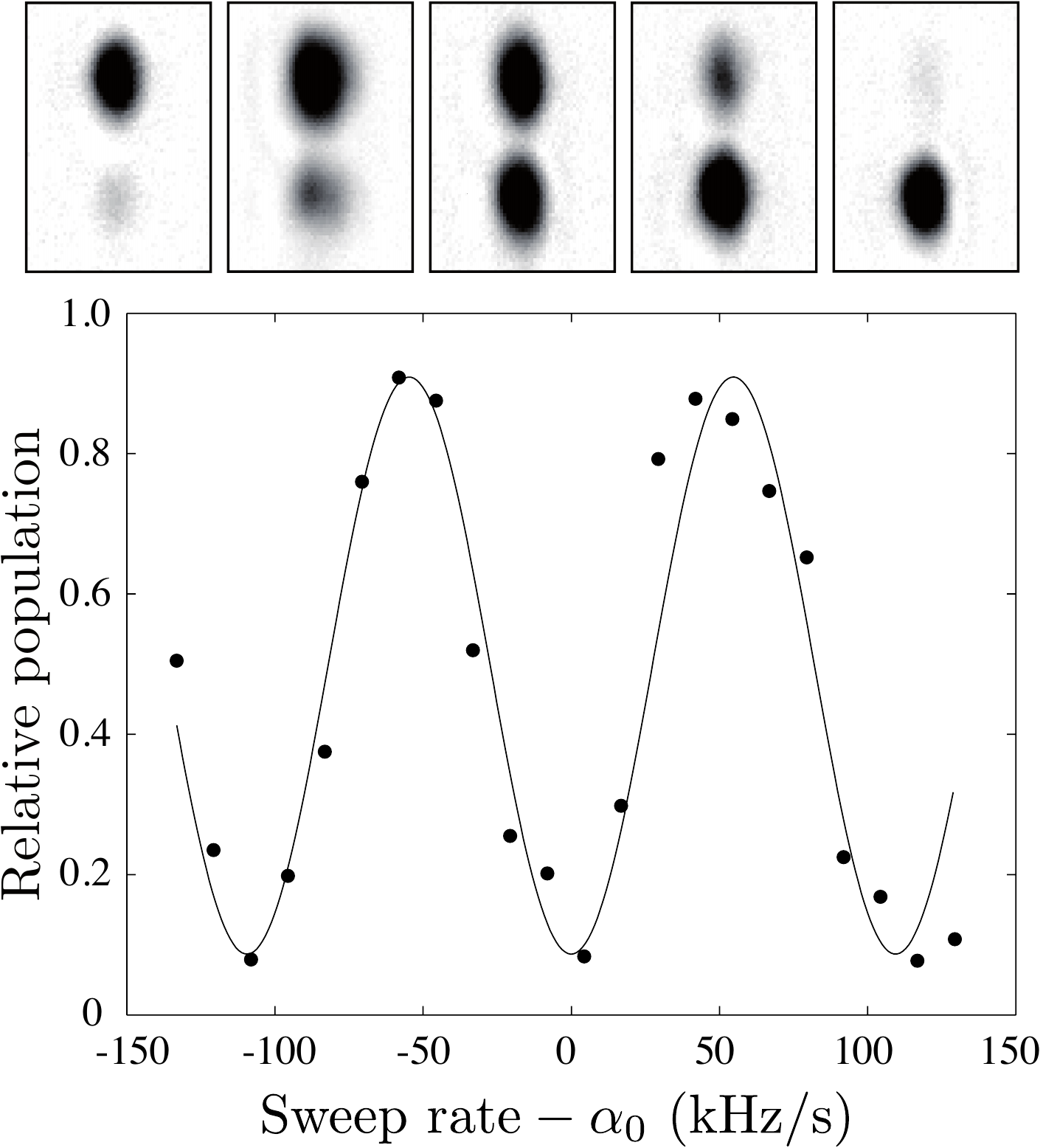}{Interference fringes from a BEC-based gravimeter with $n=1$, $T = 3$\,ms. We observe a visibility of $(83\pm6)\%$. The solid line is a least-squares sinusoidal fit to the data (as in all figures).}
%%%%

Fig. \ref{fig:highvisibility} shows gravimeter fringes for $n=1$, $T = 3$\,ms. We observe a high visibility of $(83\pm6)\%$. Increasing $T$ generally reduces the visibility, and more rapidly for larger $n$. We speculate that wavefront aberrations in the Bragg laser beams contribute to this. Aberrations cause different atomic trajectories to experience different phase shifts, as has been discussed in \cite{Fils:2005aa,Louchet-Chauvet:2011aa}. These different phase shifts are averaged through detection, causing a reduction in fringe visibility. This effect will be exacerbated for larger $n$ and $T$, as each atomic trajectory samples more of the transverse phase profile of the beam. We expect significant aberrations in the current apparatus due to the close proximity of the Bragg beams to our magnetic trapping coils, and suspect that with closer to ideal optical wavefronts, visibility would scale more weakly with $n$ and $T$. Nevertheless, we are able to improve our sensitivity by scaling the interferometer space-time area using third-order Bragg LMT beamsplitters. We achieve a mid-fringe precision of $\Delta \Phi/\Phi = 5\times 10^{-4}\text{Hz}^{-1/2}$ in this way. The corresponding fringes are given in Fig. \ref{fig:meanfield}(a). These data represent 16 minutes of acquisition time. We can determine gravity from $\alpha_0$ to be $g = 9.7859(2)$\,ms$^{-2}$. In \cite{Amalvict:2001aa}, $g$ is measured to be $9.795499189(29)$\,ms$^{-2}$, approximately 11\,km from our lab and 150\,m higher in elevation. Our value disagrees at the $10^{-2}$ level. This low accuracy is almost certainly due to the alignment of our Bragg beam along $\vec{g}$, as when calculating $g$ from $\alpha_0$, we have assumed $\vec{k}\cdot\vec{g} = kg$. We estimate an alignment uncertainty of $3^{\circ}$ in the current apparatus, which leads to a systematic error in $g$ of up to $0.026$\,ms$^{-2}$.

It is worth noting that state-of-the-art (SOA) BEC machines can produce $2.4\times10^6$ condensed atoms/s \cite{Stam:2007aa}. The best published state- and velocity-selected thermal cloud used in a gravimeter has a factor of 25 higher flux at a momentum width of $0.87\hbar k$ \cite{Muller:2008ab}. It may be the case, however, that when utilising increasingly large momentum transfer, the high spectral density of condensed sources will result in a higher usable flux due to the strict requirements on $\Delta p$, as discussed earlier. We are currently working on quantifying this effect in detail, and this will be the central topic of an upcoming paper. 

\section{THE EFFECT OF ATOMIC INTERACTIONS}

%%%%
\fig[b!]{0.58}{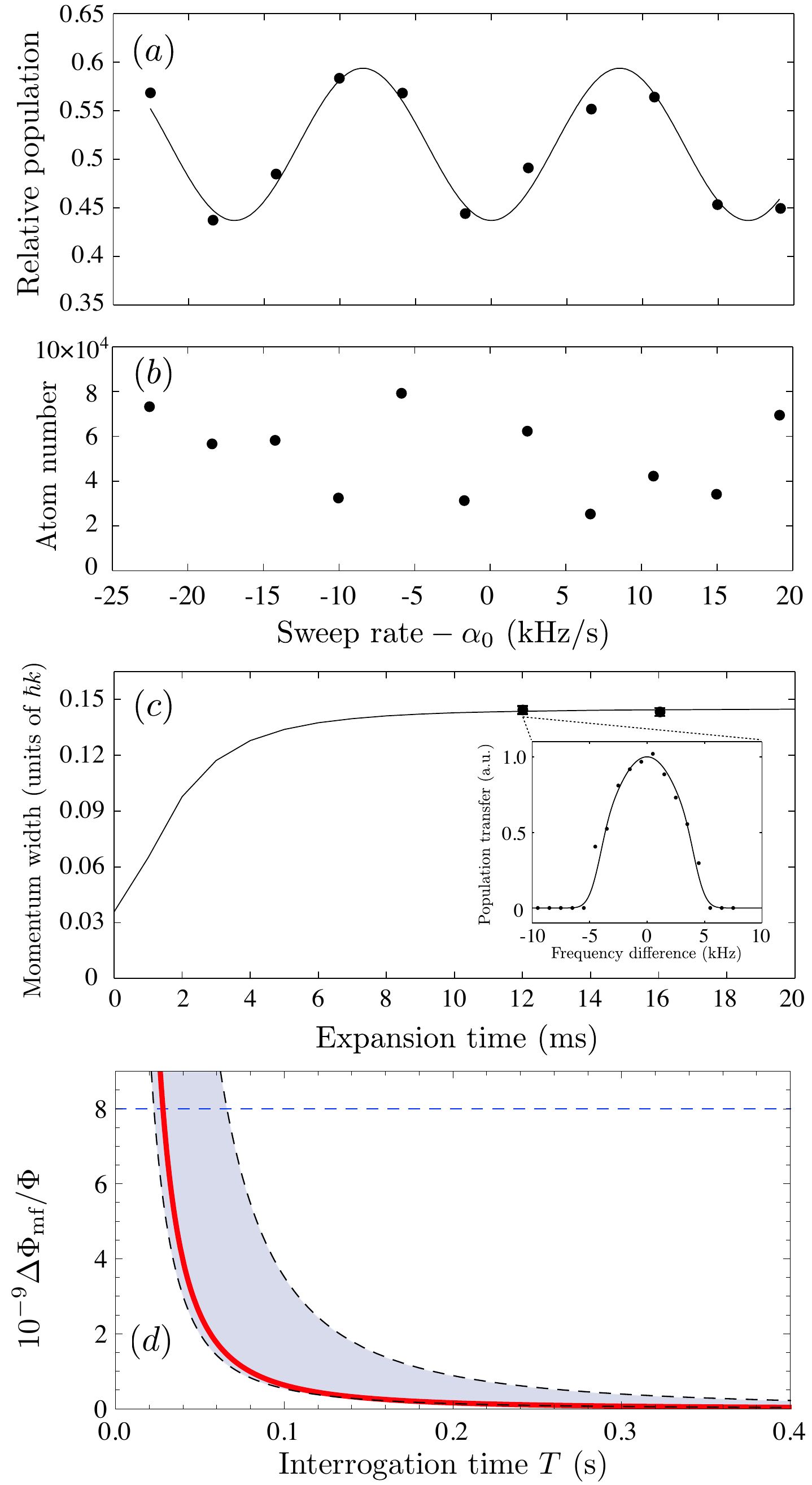}{(Color online) Density related dephasing effects. (a) Fringes for our highest precision configuration of $n=3$, $T=4$\,ms. (b) Atom number in each run of (a). (c) Vertical momentum width of the BEC as a function of expansion time. Data points are measured using Bragg spectroscopy as shown in the inset. The solid lines are from a numerical simulation of the Gross-Pitaevskii equation for our experimental setup with no free parameters. (d) Estimated dephasing limited sensitivity as a function of interrogation time. The shaded region represents values for a range of expansion times ($t_{\text{exp}}=12-40$\,ms from right to left). The solid line is the shot-noise limit for $10^6$ atoms. The dashed line is indicative of the current SOA for an atomic gravimeter (with $n=1$, $T=0.4$\,s), highlighting that for a significant range of parameters, dephasing will not limit sensitivity compared with current SOA. }
%%%%

Fig. \ref{fig:meanfield}(b) shows the atom number for each corresponding point in the fringes in Fig. \ref{fig:meanfield}(a). The variation in number at each point is intentionally imposed. Despite a variation of 300\% in density, we observe no detrimental effect of dephasing on signal-to-noise, or our measured value of $g$ at our limit of precision. After 12\,ms, the momentum width of the cloud is within 1\% of its asymptotic value, as shown in Fig. \ref{fig:meanfield}(c). We can estimate the interaction-induced phase uncertainty for our or a similar device using the following simple model. For a 50/50 beamsplitter on $N$ atoms, the variance in the number difference for the two output modes is $\sim\sqrt{N}$, giving a variance in the density of each mode. As a result of the mean-field energy shift, there will be an uncertainty in the relative phase evolution due to the energy uncertainty:
\begin{align}
E_{\text{mf}} &\simeq \frac{N}{\mathcal{V}} U \notag\\
\therefore \delta E_{\text{mf}} &\simeq \frac{\sqrt{N} U}{\mathcal{V}(t)} \simeq \frac{n(0) U }{\sqrt{N}}\frac{\mathcal{V}(0)}{ \mathcal{V}(t) }
\end{align}
where $U= 4\pi \hbar^2 a/m$ is the interaction parameter from the Gross-Pitaevskii equation, which is directly proportional to the mean-field energy, with $a$ the scattering length. The initial peak atom density is $n(0)$, and $\mathcal{V}(t)$ represents the cloud volume during expansion from a harmonic trap, which we can calculate from the Gross-Pitaevskii equation. Thus the uncertainty in the relative phase evolution is:
\begin{equation}
\omega_{\text{mf}}(t) \simeq \frac{\mu \mathcal{V}(0)}{\hbar \sqrt{N}\mathcal{V}(t)}
\end{equation}
where $\mu \simeq n(0) U$ is the chemical potential. This dephasing rate is then integrated through a given interferometer sequence to determine the phase uncertainty due to interactions, $\Delta \Phi_{\text{mf}}$. It is worth noting that as this estimate assumes a uniform initial density equal to the peak density in trap, it overestimates the effect.
 
For our current experimental parameters we find that interaction-induced dephasing would limit precision to $10^{-7}$ per shot, well below our current sensitivity. Fig. \ref{fig:meanfield}(d) projects this estimate towards SOA device parameters, plotting the dephasing-limited sensitivity as a function of $T$, for  $10^6$\, condensed atoms/s and our trap parameters. The shaded region represents the dephasing-limited sensitivity for expansion times ranging from $t_{\text{exp}} = 12-40\,$ms from the right to left boundary. The solid curve is the shot-noise limited sensitivity, and the dashed line the value of the current SOA sensitivity for an atomic gravimeter with $T=0.4$\,s and $6 \times 10^7$ atoms \cite{Muller:2008ab}. We find that with an appropriate choice of trap parameters and expansion time, interaction-induced dephasing can be made negligible compared with the shot noise limit. More importantly, as $T$ increases dephasing quickly becomes negligible compared with the current SOA precision. Very recent work has comprehensively investigated the effects of atom interactions in free-space BEC interferometers, and also confirm that interaction can be made negligible \cite{Jamison:2011aa}.

\section{COMPARING A THERMAL AND CONDENSED SOURCE}

In Fig. \ref{fig:becvsthermal}, we give a comparison of fringes using a BEC and a 100\,nK ($T/T_c \gtrsim 1$) thermal state as the source for our gravimeter, where we make every effort to ensure that the system is otherwise identical. In particular, we use an identical velocity selection pulse for each sequence. The fringe data sets were taken consecutively. The condensed source shows an improved result compared to the thermal state, with the visibility increasing from $(58\pm4)\%$ to $(85\pm11)\%$. A 500\,nK thermal state gives an even lower visibility of $(50\pm5)\%$. As discussed earlier, we speculate that wavefront aberrations contribute to the observed difference, as at 100\,nK the thermal cloud has a factor of 3 larger transverse momentum width than the condensate. Thus it will have wider range of atomic trajectories, and therefore a wider range of phase shifts across the cloud \cite{Louchet-Chauvet:2011aa}. This observation suggests that even a factor of 3 smaller transverse momentum width can improve fringe visibility if aberrations limit the interrogation time.

%%%%
\fig[h!]{0.28}{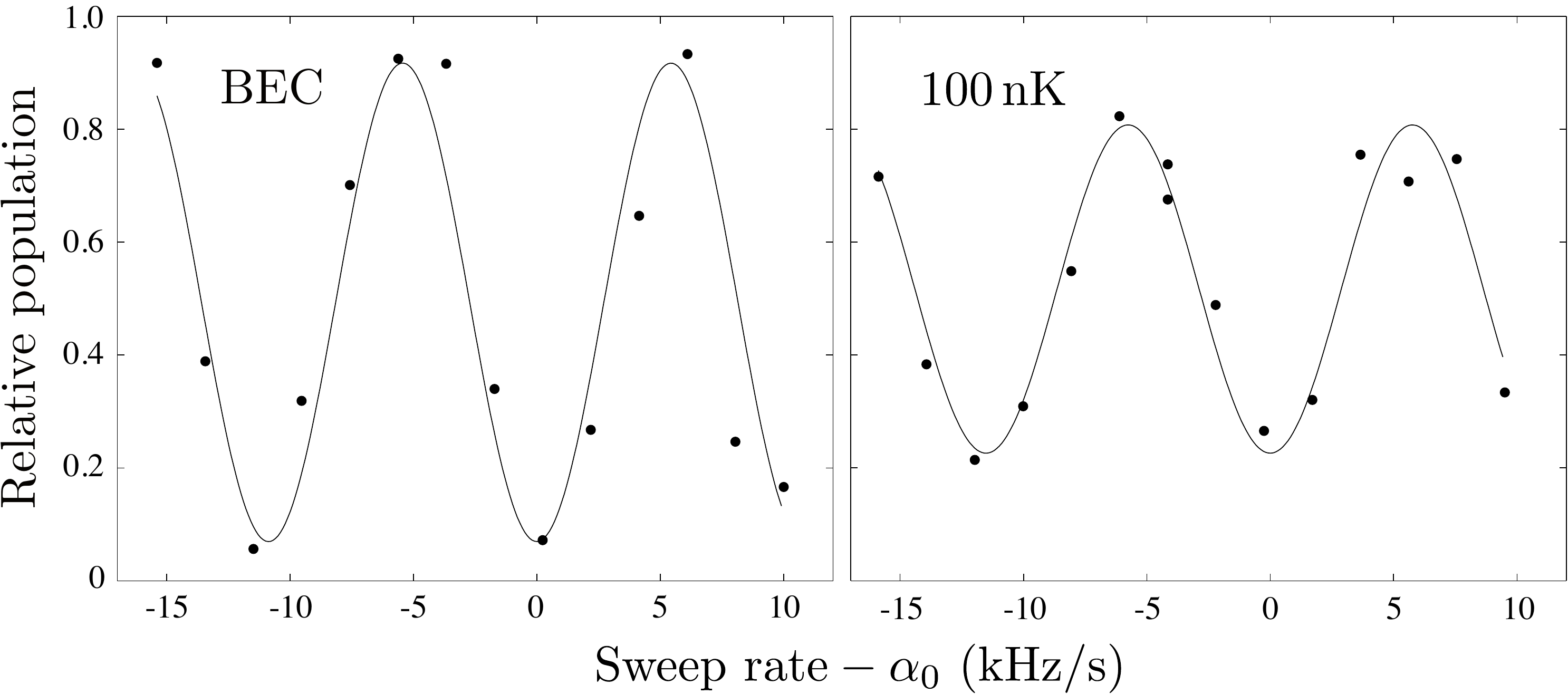}{Comparison of thermal and Bose-condensed atomic sources. We use a $n=1$, $T=3$\,ms gravimeter cycle. Both have an identical vertical momentum width, but differ in their transverse momentum width. A significant improvement in visibility from to 58\% to 85\% is seen for the BEC, with all other experimental parameters kept constant.}
%%%%

\section{BLOCH OSCILLATION BASED LMT}

We also achieve a comparatively high fringe visibility of $(24\pm4)\%$ with $6\hbar k$ LMT beamsplitters using Bloch oscillations (Fig. \ref{fig:bloch}a). After an initial $4\hbar k$ Bragg beamsplitter, we adiabatically load a lattice of depth $\sim10E_r$ with $q=0$ (stationary in the atom frame) in 100\,$\mu$s, where $q$ is the quasimomentum and $E_r$ the single-photon recoil energy. We then chirp $\delta$ over 200\,$\mu$s, sweeping $q$ through one Brillouin zone. The momentum of one arm is thus increased by $2\hbar k$ in the lab frame, as one arm remains in the lowest band whilst the other undergoes inter-band Landau-Zener transitions \cite{Clade:2009aa}. This process is reversed to decelerate this arm before the $\pi$-pulse, after which the other arm is subjected to the same procedure (Fig. \ref{fig:bloch}b). We use $T=2.5$\,ms, and our pulse sequence gives the interferometer a space-time area with an effective $n = 2.42$, in agreement with the fitted fringe period. In contrast to the work in \cite{Clade:2009aa,Muller:2009ab}, we have not used symmetric acceleration of each arm to balance differential light shifts, although our space-time area is smaller. We find that if either the lattice depth is increased beyond $10E_r$, or our acceleration time is increased to impart larger momenta, the interferometer output converges to $P=0.5$. We are currently investigating this effect further, and do not believe it to be due to randomized light-shift-induced phases as these would tend to reduce visibility without loss of contrast. We have been able to apply a $30\hbar k$ beamsplitter to our BEC with an efficiency $>95\%$, limited so far by the size of our science cell.  \\

%%%%
\fig[t]{0.62}{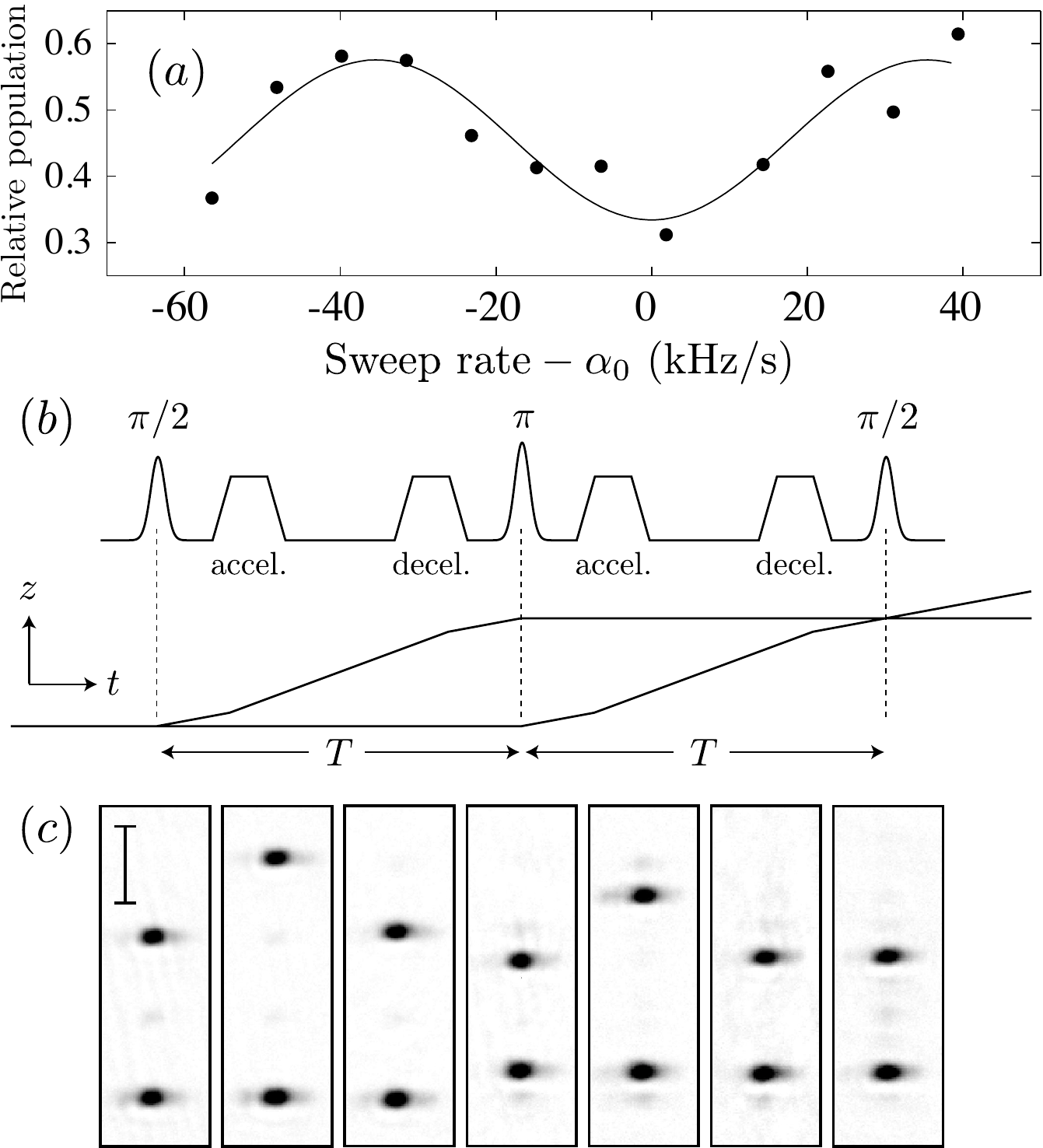}{(a) Fringes from a LMT gravimeter using Bloch beamsplitters. A visibility of 24\% is observed for $T=2.5$\,ms, and effective order $n=2.42$, calculated from the space-time area. This is in agreement with the fitted fringe period of $(70\pm5)$\,kHz. (b) The intensity of the pulse sequence used for this interferometer, and the resulting space-time diagram. Using only the Gaussian pulses results in a standard Mach-Zehnder interferometer. (c) Absorption images showing the two arms of the interferometer after each pulse. The scale bar in the (c) represents 300\,$\mu$m.}
%%%%

\section{CONCLUSIONS}

The question of whether to use coherent or thermal atomic sources for precision inertial sensing remains an important one requiring further investigation. We have presented results from the first comparison of these in a Mach-Zehnder gravimeter. We observe interference fringes with a high visibility for the condensed source, and are able to increase our sensitivity to gravity by imparting larger momentum to the atoms in the beam splitting process. The thermal source produces fringes with significantly lower visibility than the condensed source. We believe this is a result of its larger transverse momentum width, which causes higher sensitivity to wavefront aberrations in the beamsplitter lasers. We have also presented a simple model which demonstrates that interaction-induced dephasing is negligible in an interferometric measurement with freely-falling, coherent atomic samples. It may be the case that exploiting the sub-recoil momentum distribution of Bose-Einstein condensates to realise high visibility fringes with very large momentum transfer beamsplitters will lead to sensitivity beyond current state-of-the-art in precision inertial sensors.

\section*{ACKNOWLEDGMENTS} 
We would like to thank J. J. Hope, C. M. Savage, and the entire DQS atom-optics theory group for fruitful discussions. We are grateful to L. D. Turner for his support in development of the DDS control software. This work was funded by the ARC Centre of Excellence and Discovery programs.

%%%%%%%%%%%%%%%%%%%%%%%%%%%%%%%

\bibliography{grav}

\begin{thebibliography}{26}
\expandafter\ifx\csname natexlab\endcsname\relax\def\natexlab#1{#1}\fi
\expandafter\ifx\csname bibnamefont\endcsname\relax
  \def\bibnamefont#1{#1}\fi
\expandafter\ifx\csname bibfnamefont\endcsname\relax
  \def\bibfnamefont#1{#1}\fi
\expandafter\ifx\csname citenamefont\endcsname\relax
  \def\citenamefont#1{#1}\fi
\expandafter\ifx\csname url\endcsname\relax
  \def\url#1{\texttt{#1}}\fi
\expandafter\ifx\csname urlprefix\endcsname\relax\def\urlprefix{URL }\fi
\providecommand{\bibinfo}[2]{#2}
\providecommand{\eprint}[2][]{\url{#2}}

\bibitem[{\citenamefont{Kasevich~\emph{et al.}}(1992)}]{Kasevich:1992aa}
\bibinfo{author}{\bibfnamefont{M.}~\bibnamefont{Kasevich~\emph{et al.}}},
  \bibinfo{journal}{Appl. Phys. B} \textbf{\bibinfo{volume}{54}},
  \bibinfo{pages}{321} (\bibinfo{year}{1992}).

\bibitem[{\citenamefont{Canuel~\emph{et al.}}(2006)}]{Canuel:2006ab}
\bibinfo{author}{\bibfnamefont{B.}~\bibnamefont{Canuel~\emph{et al.}}},
  \bibinfo{journal}{Physical Review Letters} \textbf{\bibinfo{volume}{97}},
  \bibinfo{eid}{010402} (\bibinfo{year}{2006}).

\bibitem[{\citenamefont{M\"uller~\emph{et al.}}(2009)}]{Muller:2009aa}
\bibinfo{author}{\bibfnamefont{T.}~\bibnamefont{M\"uller~\emph{et al.}}},
  \bibinfo{journal}{Eur. Phys. J. D} \textbf{\bibinfo{volume}{53}},
  \bibinfo{pages}{273} (\bibinfo{year}{2009}).

\bibitem[{\citenamefont{Castin~\emph{et al.}}(1997)}]{Castin:1997aa}
\bibinfo{author}{\bibfnamefont{Y.}~\bibnamefont{Castin~\emph{et al.}}},
  \bibinfo{journal}{Phys. Rev. A} \textbf{\bibinfo{volume}{55}},
  \bibinfo{pages}{4330} (\bibinfo{year}{1997}).

\bibitem[{\citenamefont{Fattori~\emph{et al.}}(2008)}]{Fattori:2008aa}
\bibinfo{author}{\bibfnamefont{M.}~\bibnamefont{Fattori~\emph{et al.}}},
  \bibinfo{journal}{Phys. Rev. Lett.} \textbf{\bibinfo{volume}{100}},
  \bibinfo{eid}{080405} (\bibinfo{year}{2008}).

\bibitem[{\citenamefont{Peters~\emph{et al.}}(2001)}]{Peters:2001aa}
\bibinfo{author}{\bibfnamefont{A.}~\bibnamefont{Peters~\emph{et al.}}},
  \bibinfo{journal}{Metrologia} \textbf{\bibinfo{volume}{38}},
  \bibinfo{pages}{25} (\bibinfo{year}{2001}).

\bibitem[{\citenamefont{M\"uller~\emph{et
  al.}}(2008{\natexlab{a}})}]{Muller:2008ab}
\bibinfo{author}{\bibfnamefont{H.}~\bibnamefont{M\"uller~\emph{et al.}}},
  \bibinfo{journal}{Phys. Rev. Lett.} \textbf{\bibinfo{volume}{100}},
  \bibinfo{pages}{031101} (\bibinfo{year}{2008}{\natexlab{a}}).

\bibitem[{\citenamefont{Denschlag~\emph{et al.}}(2002)}]{Denschlag:2002aa}
\bibinfo{author}{\bibfnamefont{J.~H.} \bibnamefont{Denschlag~\emph{et al.}}},
  \bibinfo{journal}{J. Phys. B} \textbf{\bibinfo{volume}{35}},
  \bibinfo{pages}{3095} (\bibinfo{year}{2002}).

\bibitem[{\citenamefont{M\"uller~\emph{et
  al.}}(2008{\natexlab{b}})}]{Muller:2008aa}
\bibinfo{author}{\bibfnamefont{H.}~\bibnamefont{M\"uller~\emph{et al.}}},
  \bibinfo{journal}{Phys. Rev. Lett.} \textbf{\bibinfo{volume}{100}},
  \bibinfo{eid}{180405} (\bibinfo{year}{2008}{\natexlab{b}}).

\bibitem[{\citenamefont{M\"uller et~al.}(2009)\citenamefont{M\"uller, Chiow,
  Herrmann, and Chu}}]{Muller:2009ab}
\bibinfo{author}{\bibfnamefont{H.}~\bibnamefont{M\"uller}},
  \bibinfo{author}{\bibfnamefont{S.-w.} \bibnamefont{Chiow}},
  \bibinfo{author}{\bibfnamefont{S.}~\bibnamefont{Herrmann}}, \bibnamefont{and}
  \bibinfo{author}{\bibfnamefont{S.}~\bibnamefont{Chu}},
  \bibinfo{journal}{Phys. Rev. Lett.} \textbf{\bibinfo{volume}{102}},
  \bibinfo{pages}{240403} (\bibinfo{year}{2009}).

\bibitem[{\citenamefont{Clad\'e~\emph{et al.}}(2009)}]{Clade:2009aa}
\bibinfo{author}{\bibfnamefont{P.}~\bibnamefont{Clad\'e~\emph{et al.}}},
  \bibinfo{journal}{Phys. Rev. Lett.} \textbf{\bibinfo{volume}{102}},
  \bibinfo{pages}{240402} (\bibinfo{year}{2009}).

\bibitem[{\citenamefont{Giltner~\emph{et al.}}(1995)}]{Giltner:1995aa}
\bibinfo{author}{\bibfnamefont{D.~M.} \bibnamefont{Giltner~\emph{et al.}}},
  \bibinfo{journal}{Phys. Rev. Lett.} \textbf{\bibinfo{volume}{75}},
  \bibinfo{pages}{2638} (\bibinfo{year}{1995}).

\bibitem[{\citenamefont{Kozuma~\emph{et al.}}(1999)}]{Kozuma:1999ab}
\bibinfo{author}{\bibfnamefont{M.}~\bibnamefont{Kozuma~\emph{et al.}}},
  \bibinfo{journal}{Phys. Rev. Lett.} \textbf{\bibinfo{volume}{82}},
  \bibinfo{pages}{871} (\bibinfo{year}{1999}).

\bibitem[{not()}]{note2}
\bibinfo{note}{We define visibility through the ability to fit a sine curve to
  a given data set. Contrast is indicative of coherence, and it is possible to
  have contrast with no visibility, as is discussed in
  \cite{Chiow:2009aa,Muller:2008aa}.}

\bibitem[{\citenamefont{Torii~\emph{et al.}}(2000)}]{Torii:2000aa}
\bibinfo{author}{\bibfnamefont{Y.}~\bibnamefont{Torii~\emph{et al.}}},
  \bibinfo{journal}{Phys. Rev. A} \textbf{\bibinfo{volume}{61}},
  \bibinfo{pages}{041602} (\bibinfo{year}{2000}).

\bibitem[{\citenamefont{Le~Gou{\"e}t~\emph{et al.}}(2008)}]{Le-Gouet:2008aa}
\bibinfo{author}{\bibfnamefont{J.}~\bibnamefont{Le~Gou{\"e}t~\emph{et al.}}},
  \bibinfo{journal}{Appl. Phys. B} \textbf{\bibinfo{volume}{92}},
  \bibinfo{pages}{133} (\bibinfo{year}{2008}).

\bibitem[{\citenamefont{Hughes~\emph{et al.}}(2009)}]{Hughes:2009aa}
\bibinfo{author}{\bibfnamefont{K.~J.} \bibnamefont{Hughes~\emph{et al.}}},
  \bibinfo{journal}{Phys. Rev. Lett.} \textbf{\bibinfo{volume}{102}},
  \bibinfo{pages}{150403} (\bibinfo{year}{2009}).

\bibitem[{\citenamefont{Stenger~\emph{et al.}}(1999)}]{Stenger:1999aa}
\bibinfo{author}{\bibfnamefont{J.}~\bibnamefont{Stenger~\emph{et al.}}},
  \bibinfo{journal}{Phys. Rev. Lett.} \textbf{\bibinfo{volume}{82}},
  \bibinfo{pages}{4569} (\bibinfo{year}{1999}).

\bibitem[{\citenamefont{Altin~\emph{et al.}}(2010)}]{Altin:2010ab}
\bibinfo{author}{\bibfnamefont{P.~A.} \bibnamefont{Altin~\emph{et al.}}},
  \bibinfo{journal}{Rev. Sci. Instrum.} \textbf{\bibinfo{volume}{81}},
  \bibinfo{eid}{063103} (\bibinfo{year}{2010}).

\bibitem[{\citenamefont{M\"uller~\emph{et
  al.}}(2008{\natexlab{c}})}]{Muller:2008ac}
\bibinfo{author}{\bibfnamefont{H.}~\bibnamefont{M\"uller~\emph{et al.}}},
  \bibinfo{journal}{Phys. Rev. A} \textbf{\bibinfo{volume}{77}},
  \bibinfo{pages}{023609} (\bibinfo{year}{2008}{\natexlab{c}}).

\bibitem[{\citenamefont{Amalvict~\emph{et al.}}(2001)}]{Amalvict:2001aa}
\bibinfo{author}{\bibfnamefont{M.}~\bibnamefont{Amalvict~\emph{et al.}}},
  \bibinfo{journal}{J. Geodet. Soc. Jpn.} \textbf{\bibinfo{volume}{47}},
  \bibinfo{pages}{410} (\bibinfo{year}{2001}).

\bibitem[{\citenamefont{Fils~\emph{et al.}}(2005)}]{Fils:2005aa}
\bibinfo{author}{\bibfnamefont{J.}~\bibnamefont{Fils~\emph{et al.}}},
  \bibinfo{journal}{The European Physical Journal D-Atomic, Molecular and
  Optical Physics} \textbf{\bibinfo{volume}{36}}, \bibinfo{pages}{257}
  (\bibinfo{year}{2005}).

\bibitem[{\citenamefont{Louchet-Chauvet
  et~al.}(2011)\citenamefont{Louchet-Chauvet, Farah, Bodart, Clairon,
  Landragin, Merlet, and Santos}}]{Louchet-Chauvet:2011aa}
\bibinfo{author}{\bibfnamefont{A.}~\bibnamefont{Louchet-Chauvet}},
  \bibinfo{author}{\bibfnamefont{T.}~\bibnamefont{Farah}},
  \bibinfo{author}{\bibfnamefont{Q.}~\bibnamefont{Bodart}},
  \bibinfo{author}{\bibfnamefont{A.}~\bibnamefont{Clairon}},
  \bibinfo{author}{\bibfnamefont{A.}~\bibnamefont{Landragin}},
  \bibinfo{author}{\bibfnamefont{S.}~\bibnamefont{Merlet}}, \bibnamefont{and}
  \bibinfo{author}{\bibfnamefont{F.~P.~D.} \bibnamefont{Santos}},
  \bibinfo{journal}{New Journal of Physics} \textbf{\bibinfo{volume}{13}},
  \bibinfo{pages}{065025} (\bibinfo{year}{2011}),
  \urlprefix\url{http://stacks.iop.org/1367-2630/13/i=6/a=065025}.

\bibitem[{\citenamefont{van~der Stam~\emph{et al.}}(2007)}]{Stam:2007aa}
\bibinfo{author}{\bibfnamefont{K.~M.~R.} \bibnamefont{van~der Stam~\emph{et
  al.}}}, \bibinfo{journal}{Review of Scientific Instruments}
  \textbf{\bibinfo{volume}{78}}, \bibinfo{eid}{013102}
  (pages~\bibinfo{numpages}{10}) (\bibinfo{year}{2007}).

\bibitem[{\citenamefont{{Jamison} et~al.}(2011)\citenamefont{{Jamison}, {Kutz},
  and {Gupta}}}]{Jamison:2011aa}
\bibinfo{author}{\bibfnamefont{A.~O.} \bibnamefont{{Jamison}}},
  \bibinfo{author}{\bibfnamefont{J.~N.} \bibnamefont{{Kutz}}},
  \bibnamefont{and} \bibinfo{author}{\bibfnamefont{S.}~\bibnamefont{{Gupta}}},
  \bibinfo{journal}{Arxiv:1103.1454}  (\bibinfo{year}{2011}).

\bibitem[{\citenamefont{Chiow~\emph{et al.}}(2009)}]{Chiow:2009aa}
\bibinfo{author}{\bibfnamefont{S.-w.} \bibnamefont{Chiow~\emph{et al.}}},
  \bibinfo{journal}{Phys. Rev. Lett.} \textbf{\bibinfo{volume}{103}},
  \bibinfo{pages}{050402} (\bibinfo{year}{2009}).

\end{thebibliography}

%%%%%%%%%%%%%%%%%%%%%%%%%%%%%%%

\end{document}